\documentclass{aa}

\usepackage[varg]{txfonts}
\usepackage{siunitx}
\usepackage[version=4]{mhchem}
\usepackage{multirow}
\usepackage{graphicx}
\usepackage[caption=false]{subfig}
\usepackage[strict]{changepage}
\usepackage{xcolor}
\usepackage{ulem}
\usepackage[colorlinks=true,citecolor=blue]{hyperref}
\usepackage{float}
\usepackage{amssymb}
\usepackage{amsmath}
\usepackage{lscape}
\usepackage{mathtools}
\usepackage{natbib}
\usepackage{pdflscape}
\usepackage{gensymb}

\defcitealias{zhou2023}{Paper~I}   

\DeclareSIUnit\jansky{Jy}

\newcolumntype{L}[1]{>{\raggedright\let\newline\\\arraybackslash\hspace{0pt}}m{#1}}
\newcolumntype{C}[1]{>{\centering\let\newline\\\arraybackslash\hspace{0pt}}m{#1}}
\newcolumntype{R}[1]{>{\raggedleft\let\newline\\\arraybackslash\hspace{0pt}}m{#1}}

\begin{document}

\title{The currently observed clumps cannot be the "direct" precursors of the currently observed open clusters}

\author{J. W. Zhou\inst{\ref{inst1}} 
\and Sami Dib \inst{\ref{inst2}}
\and Pavel Kroupa\inst{\ref{inst3}} \fnmsep \inst{\ref{inst4}}
}
\institute{
Max-Planck-Institut f\"{u}r Radioastronomie, Auf dem H\"{u}gel 69, 53121 Bonn, Germany \label{inst1} \\
\email{jwzhou@mpifr-bonn.mpg.de}
\and
Max Planck Institute f\"{u}r Astronomie, K\"{o}nigstuhl 17, 69117 Heidelberg, Germany \label{inst2}\\
\email{sami.dib@gmail.com}
\and
Helmholtz-Institut f{\"u}r Strahlen- und Kernphysik (HISKP), Universität Bonn, Nussallee 14–16, 53115 Bonn, Germany \label{inst3}\\
\email{pkroupa@uni-bonn.de}
\and
Charles University in Prague, Faculty of Mathematics and Physics, Astronomical Institute, V Hole{\v s}ovi{\v c}k{\'a}ch 2, CZ-180 00 Praha 8, Czech Republic \label{inst4}
}
\date{Accepted XXX. Received YYY; in original form ZZZ}

\abstract
{We categorized clumps, embedded clusters, and open clusters and conducted a comparative analysis of their physical properties. 
Overall, the radii of open clusters are significantly larger than those of embedded clusters and clumps. The radii of embedded clusters are larger than those of clumps, which may be due to the expansion of embedded clusters.
The open clusters have significantly higher masses than embedded clusters, by about one order of magnitude.
Given the current mass distribution of clumps in the Milky Way, the evolutionary sequence from a single clump evolving into an embedded cluster and subsequently into an open cluster cannot account for the observed open clusters with old ages and high masses, which is also supported by N-body simulations of individual embedded clusters. 
To explain the mass and radius distributions of the observed open clusters, initial embedded clusters with masses higher than 3000 M$_{\odot}$ are necessary. 
However, the upper limit of the embedded cluster sample is less than 1000 M$_{\odot}$, and only a few ATLASGAL clumps have a mass higher than 3000 M$_{\odot}$.
Thus, the currently observed clumps cannot be the "direct" precursors of the currently observed open clusters. 
If the Milky Way has a burst-like and time-dependent star formation history, the currently observed open clusters with old ages and high masses may come from massive clumps in the past. There is also a very real possibility that these open clusters originate from post-gas expulsion coalescence of multiple embedded clusters. We compared  the separation of open clusters and the typical size of molecular clouds, and find that most molecular clouds may only form one open cluster, which supports the scenario of post-gas expulsion coalescence. Further study is necessary to distinguish between the different scenarios.}



\titlerunning{}
\authorrunning{}

\maketitle 

\section{Introduction}

Massive star-forming regions (MSFRs) constitute a primary mode of star formation within the Galaxy \citep{Lada2003-41, Motte2018-56}. Young stellar clusters within these regions can potentially evolve into open clusters (OCs). It has been proposed that the majority of observed stars, if not all, originate from embedded clusters \citep{Kroupa1995a-277, Kroupa1995b-277, Lada2003-41, Kroupa2005-576, Megeath2016-151, Dinnbier2022-660}. Despite their significance, the characteristics of embedded clusters and their connection to OCs remain unclear. The physical processes responsible for the fragmentation of cluster-forming gas and the subsequent dynamics of the resulting stars are not well understood \citep{Megeath2016-151}. While it is probable that embedded clusters are precursors to OCs, \citet{Lada2003-41} found that only 7\% of embedded clusters survive the gas dispersal phase.

Understanding the structure and dynamical evolution of embedded clusters represents a crucial step toward unraveling the formation processes of OCs. Recent observational advancements indicate that early expansion plays a pivotal role in shaping the development of young star clusters. Utilizing {\it Gaia} Data Release (DR) 2 and DR3 data alongside multi-object spectroscopy, studies have unveiled the expansion phenomena in very young clusters within star-forming regions through both case studies \citep{Wright2019-486, Cantat2019-626, Kuhn2020-899, Lim2020-899, Swiggum2021-917, Lim2022-163, Muzic2022-668, Das2023-948, Flaccomio2023-670} and statistical works \citep{Kuhn2019-870, Della2023arXiv, Wright2023arXiv}.
N-body simulations of massive star clusters with monolithic structures and residual gas expulsion \citep{Kroupa2001-321, Banerjee2013-764, Banerjee2014-787, Banerjee2015-447,Haghi2020-904} as well as semi-analytical models \citep{Dib2007-381,Dib2010-405,Dib2023-959} have consistently replicated the characteristics of well-observed very young massive clusters such as the Orion Nebula Cluster, R136, NGC 3603, and the Arches cluster, offering insights into their initial formation conditions. For lower-mass embedded clusters, \citet{Zhou2024rm} compiled various cluster samples to examine the early expansion dynamics using numerical simulation recipes from \citet{Kroupa2001-321}, \citet{Banerjee2013-764, Banerjee2014-787, Banerjee2015-447}, and \citet{Dinnbier2022-660}. A special focus of that work was  the evolution of the initial mass-radius relation of embedded clusters.

In \citet{Zhou2024rm}, direct N-body simulations were employed to model the early expansion of embedded clusters after the expulsion of their residual gas. The models effectively fit the observational data on cluster radii from various sources, indicating that observed very young clusters are likely in an expanding phase. They also suggest that even embedded clusters within ATLASGAL clumps with HII regions may already be undergoing expansion.
In \citet{Zhou2024-688}, the dendrogram algorithm was employed to analyze surface density maps of stars derived from the Massive Young Star-Forming Complex Study in Infrared and X-ray (MYStIX) project \citep{Feigelson2013-209, Kuhn2015-802}. This approach decomposes the maps into hierarchical structures, revealing the multi-scale features of star clusters.
Additionally, the minimum spanning tree (MST) method was utilized to quantify the distances between clusters and clumps within each MSFR. Notably, the distances between clusters, between clumps, and between clusters and clumps were found to be comparable. This suggests that the evolution from clumps to embedded clusters occurs locally and independently, with minimal influence on surrounding objects.
In this work we further investigated the evolution from embedded clusters to OCs by analyzing observational data and conducting N-body simulations. 

\section{Sample}

\subsection{Embedded clusters}

In \citet{Zhou2024-688} we identified a sample of embedded clusters based on the surface density maps of stars for 17 MSFRs with distances less than 3.6~kpc from the MYStIX project. To this we added the embedded cluster catalogs of \citet{Lada1991-371}, \citet{Carpenter1993-407}, \citet{Lada2003-41}, \citet{Carpenter2000-120}, \citet{Kumar2006-449}, and \citet{Faustini2009-503} and required the distance of the clusters to be less than 3.6~kpc.

\subsection{Clumps}

We selected clumps from the ATLASGAL survey \citep{Urquhart2022-510} and limited the kinematic distance of the clumps to be less than 3.6~kpc.
In addition to being consistent with the distance restriction of the embedded cluster sample, this distance limitation can also effectively eliminate a possible distance bias \citep{Urquhart2022-510}.

\subsection{Open clusters}
\citet{Hunt2023-673}
conducted a blind, all-sky search for OCs using 729 million sources from {\it Gaia} DR3 down to magnitude G $\sim$20, creating a homogeneous catalog of 7167 clusters. The \citet{Hunt2024-686} catalog (see their Table 3) includes 5647 OCs and 1309 moving groups, of which 3530 OCs and 539 moving groups are of high quality. For the 3530 high-quality OCs, most of them (3103) have a distance of less than 3.6~kpc. 

\section{Results}

\subsection{Mass and radius}

\begin{figure}[htbp!]
\centering
\includegraphics[width=0.45\textwidth]{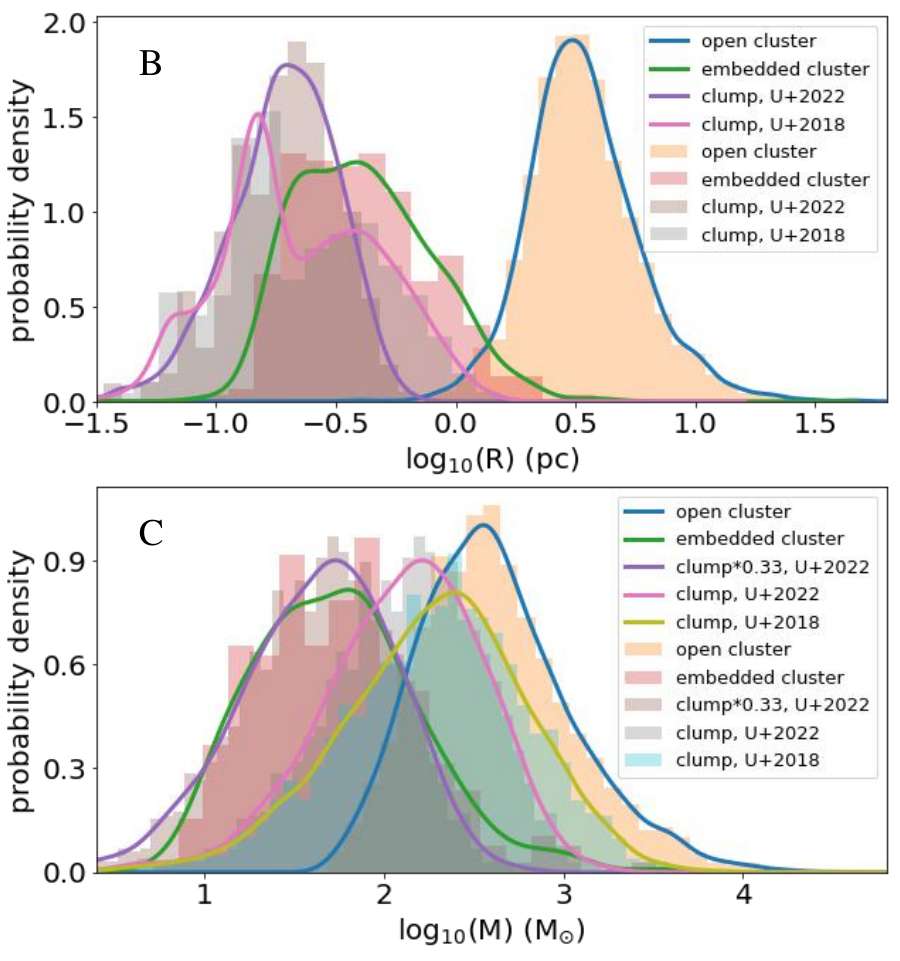}
\caption{Physical properties of Galactic clumps, embedded clusters, and open clusters.
(a) Radius distribution; (b) Mass distribution. The lines are from the histograms.}
\label{compare}
\end{figure}

For the ATLASGAL clumps in \citet{Urquhart2022-510}, the radius was determined from the number of pixels within the full width at half maximum (FWHM) contour (i.e., above 50 percent of the peak of the ATLASGAL dust continuum emission). The clump's mass, also measured at the FWHM, was determined by integrating the flux density at 870 $\mu$m within the 50 percent contour. Furthermore, the radii of ATLASGAL clumps with HII regions (HII clumps) were confirmed to be equal to the half-mass radii of embedded clusters in the clumps in \citet{Zhou2024rm}.
In \citet{Hunt2024-686}, the radius of OCs was determined by including half of the members (r$_{\rm 50}$) that is close to the half-mass radius. 
For embedded clusters, their radii were unified to the half-mass radii, as done in \citet{Zhou2024rm}. 
Therefore, the radii of clumps, embedded clusters, and OCs can be directly compared. Generally, the radii of OCs are significantly larger than those of embedded clusters and clumps.
The radii of embedded clusters are larger than those of clumps, which may be due to the expansion of embedded clusters, as presented in \citet{Zhou2024rm} and clearly shown in Appendix \ref{msfr}.

In Fig.\ref{compare}(b), the mass distribution decreases from OCs to clumps to embedded clusters. Interestingly, the mass distribution of clumps, when multiplied by 0.33, is comparable to that of embedded clusters, implying that 0.33 could be a typical star formation efficiency (SFE) for clumps in the Milky Way. 
The mass distribution of the clumps is significantly smaller than that of the OCs. The mass of ATLASGAL clumps in \citet{Urquhart2018-473} is the full mass rather than the FWHM mass in \citet{Urquhart2022-510}, which is still significantly lower than the mass of OCs.

These offsets raise questions about the origin of OCs with masses significantly higher than those of their progenitors (clumps or embedded clusters).
From clumps to embedded clusters,
the SFE is $\approx$0.3 \citep{Lada2003-41,Megeath2016-151,Zhou2024-688,Zhou2024arXiv240809867Z,zhou2024starformationhistoriesstar}, and there is considerable mass loss during the evolution. Therefore, for a single clump to evolve into a single embedded cluster and then evolve into a single OC, with the current clump mass distribution in the Milky Way, this evolutionary sequence cannot produce the observed OCs' mass distribution, as verified by the direct N-body simulations in Sect. \ref{single}.

If an OC originates from an embedded cluster, the initial mass of the embedded cluster should be much higher than that of the OC due to the considerable mass loss during the evolution, as shown in Sect. \ref{ana}.
However, the opposite is seen
in Fig.\ref{compare}(b), where the mass distribution of OCs is significantly larger than that of embedded clusters (by about one order of magnitude), suggesting that most OCs do not originate from single embedded clusters.

\subsection{Separations of open clusters}\label{sepa}

\begin{figure}[htbp!]
\centering
\includegraphics[width=0.48\textwidth]{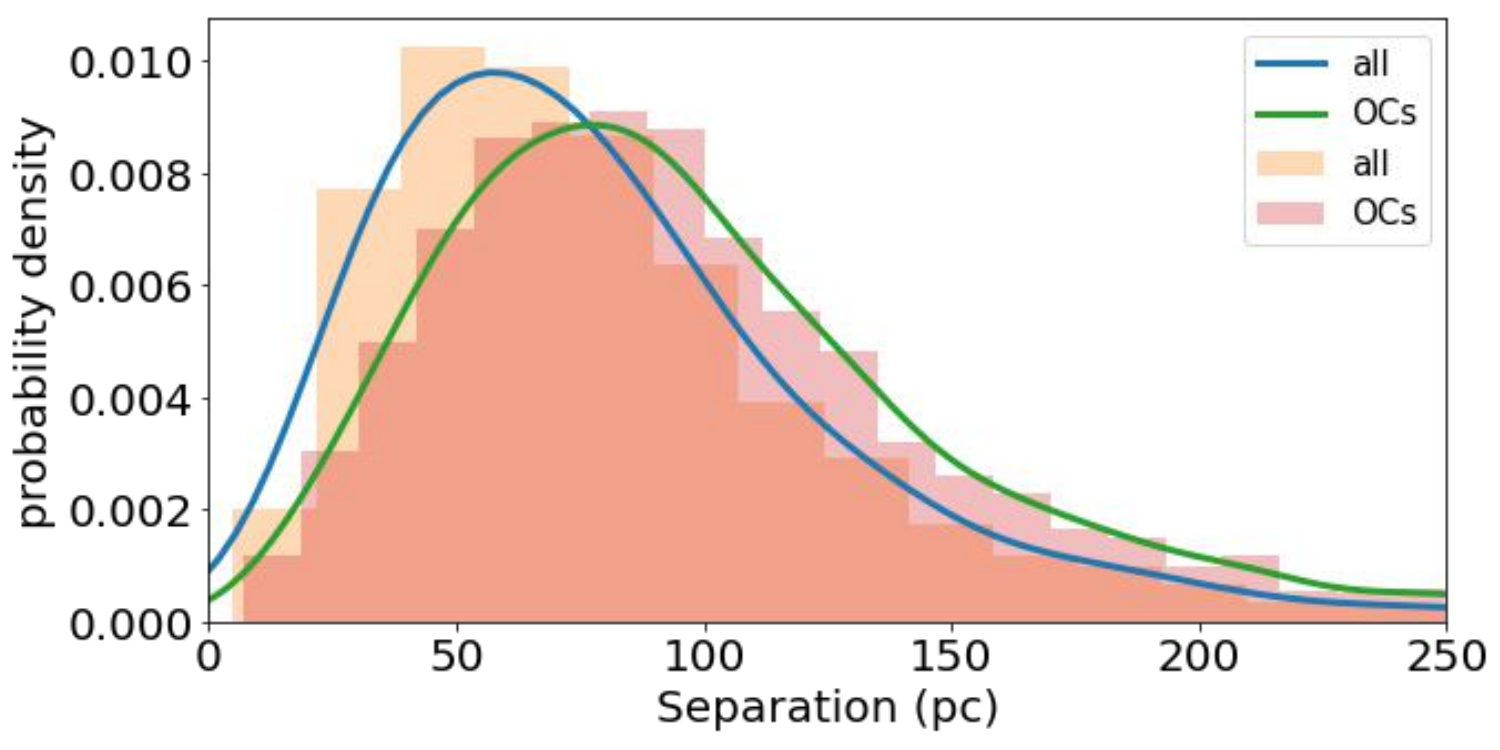}
\caption{Separation distribution of each of the two neighbouring objects. The "all" signifies all 5858 OCs and moving groups, and "OCs" high-quality OCs.}
\label{mst}
\end{figure}

We used the minimum spanning tree (MST) analysis to measure the inter-cluster spacing based on the 3D coordinates of OCs. The MST connects all clusters in a network with the least amount of total distance, without creating any loops. The length of the line connecting two clusters represents the separation between them.
In Fig. \ref{mst}  the typical inter-cluster separation of 3103 high-quality OCs is $\sim$80 pc. Considering all 5858 OCs and moving groups (< 3.6~kpc), the lower limit of the separation is $\sim$60 pc. 
However, the most probable size for molecular clouds in the entire Milky Way disk is $\sim$30 pc \citep{Miville2017-834}.
This implies that most molecular clouds  may only form one OC. 
Each molecular cloud typically includes many embedded clusters, as presented in Appendix \ref{msfr}. If each individual embedded cluster could evolve into an OC, the spacing between OCs should not be so large; it should at least be smaller than the size of the parental molecular cloud.

In \citet{Zhou2024-688}, for the MSFRs in the MYStIX project,
the separations between clusters, between clumps, and between clusters and clumps are quite comparable. The typical separation is $\sim$1 pc. Therefore, as the embedded clusters in each MSFR expand, they are able to undergo merger (post-gas expulsion coalescence). Significant differences in separation for embedded clusters and OCs indicate that there are considerable mergers between embedded clusters as they evolve into OCs. We will discuss this picture in a forthcoming work.

\subsection{Single embedded cluster simulation}\label{single}

\begin{figure}[htbp!]
\centering
\includegraphics[width=0.48\textwidth]{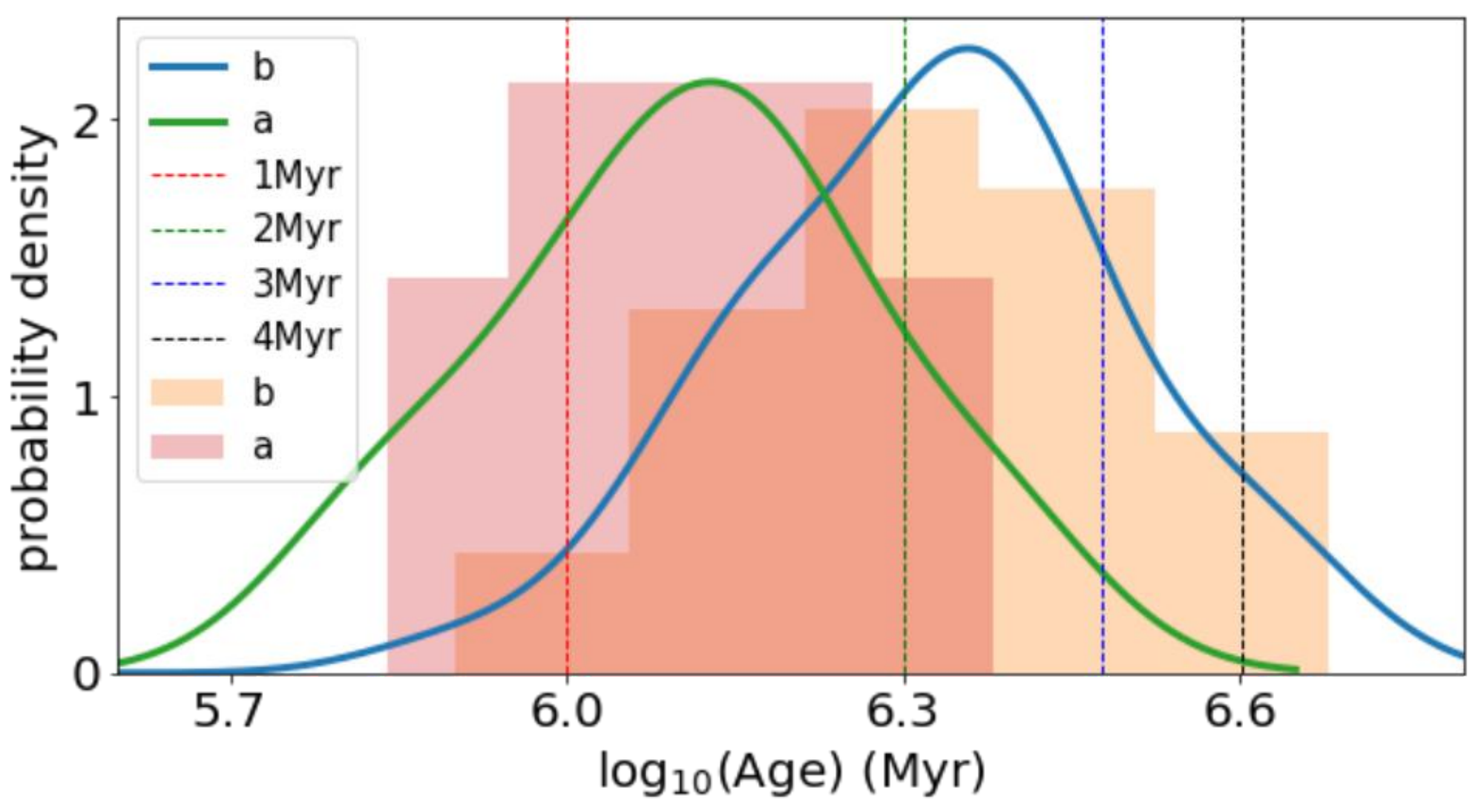}
\caption{Age distribution of embedded clusters in the seven MSFRs displayed in Appendix \ref{msfr}. The ages are taken from Table 1 of \citet{Kuhn2015-812}. The "a" represents NGC~6334 and M~17, and "b" represents the Lagoon nebula, NGC~6357, the Eagle nebula, the Carina nebula, and the Trifid nebula.
}
\label{age}
\end{figure}

\begin{figure}[htbp!]
\centering
\includegraphics[width=0.48\textwidth]{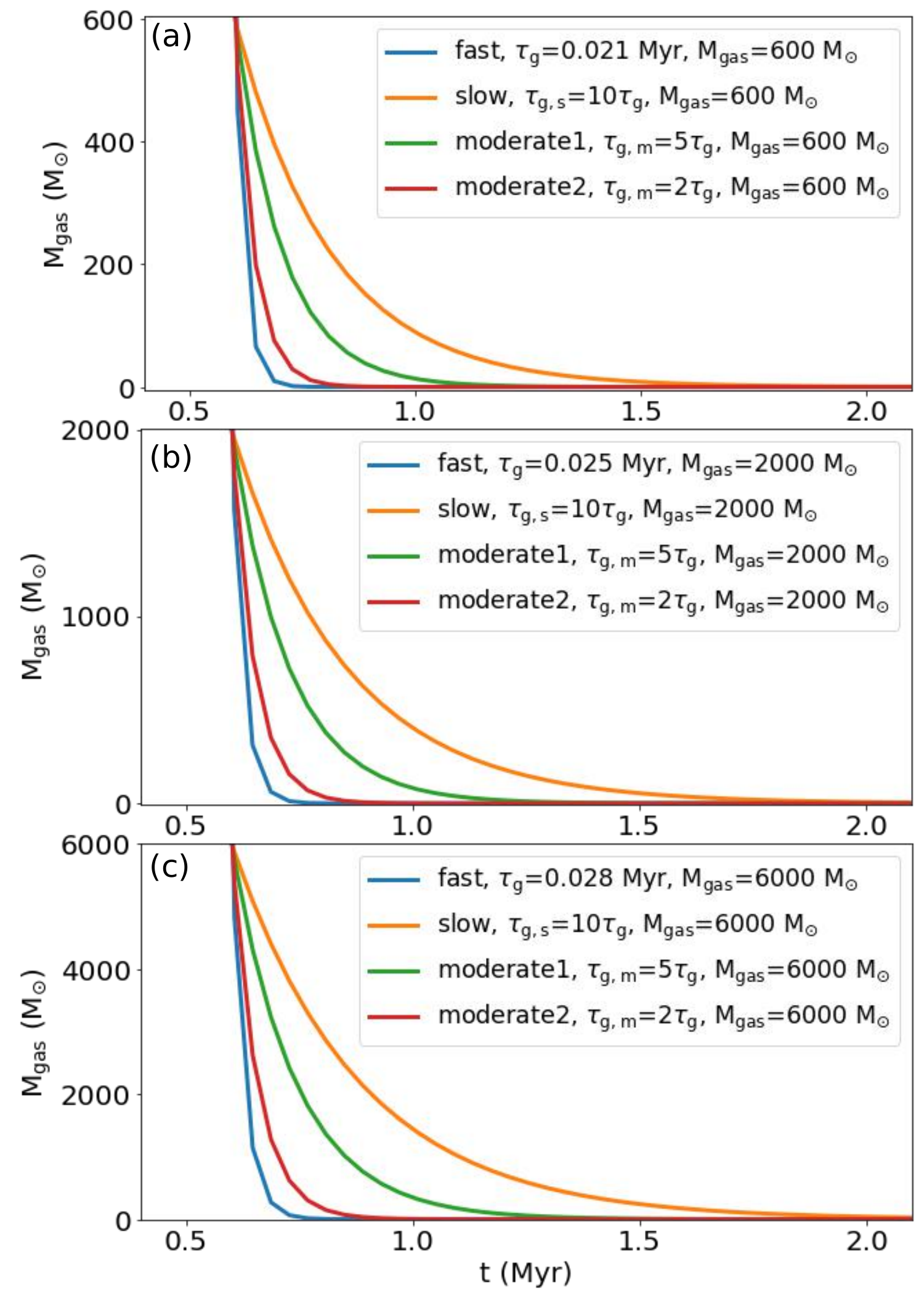}
\caption{Evolution of the gas mass over time under different gas expulsion modes.}
\label{gas}
\end{figure}

\begin{figure}[htbp!]
\centering
\includegraphics[width=0.48\textwidth]{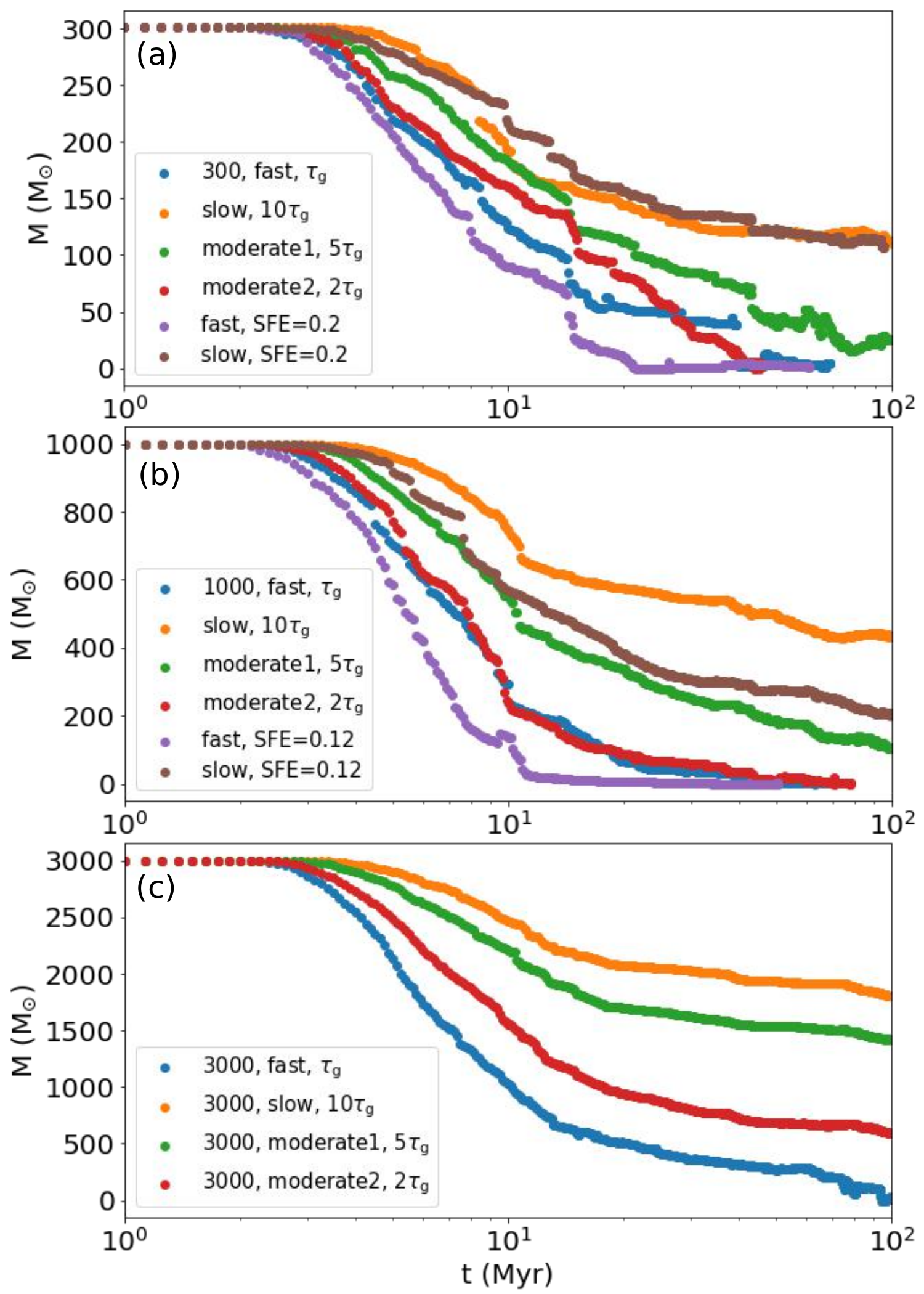}
\caption{Evolution of the cluster mass within the tidal radius over time under different gas expulsion modes and with different SFEs.}
\label{simu-s}
\end{figure}

\begin{figure*}[htbp!]
\centering
\includegraphics[width=1\textwidth]{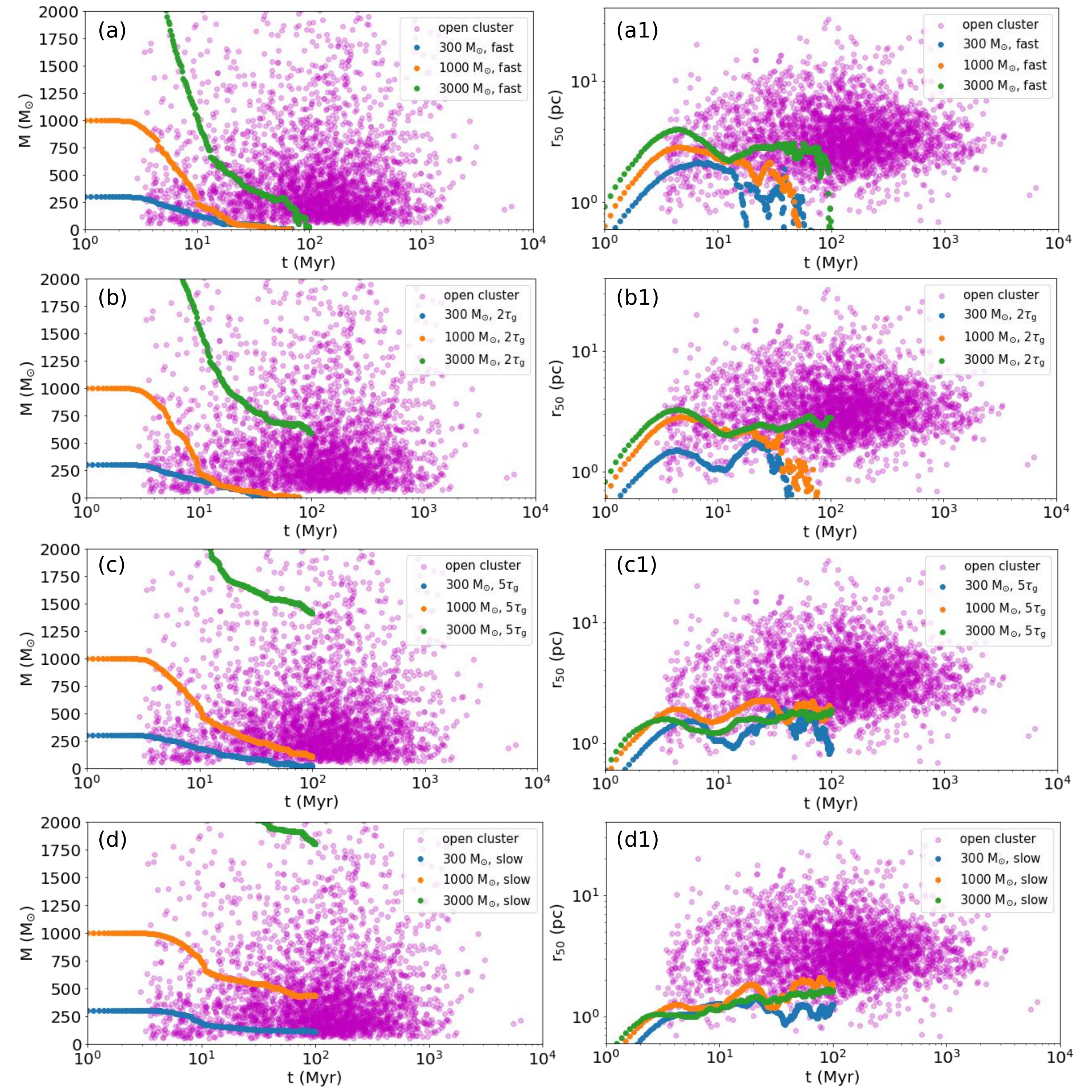}
\caption{Fitting the physical parameters of OCs in \citet{Hunt2024-686} using single embedded cluster simulations.}
\label{fit0}
\end{figure*}

\begin{table*}
\centering
\caption{Computed model parameters. 
}
\label{tab1}
\begin{tabular}{ccccccccc}
\hline
        $M_{\rm ecl} ({\rm M_{\odot}})$ &       $r_{\rm h} ({\rm pc})$  &       $m_{\rm max} ({\rm M_{\odot}})$ &       $\tau_{\rm g} ({\rm Myr})$ & $\tau_{\rm d} ({\rm Myr})$&-s (Myr)&-r (pc)&--r-search-min(pc)&--r-bin(pc)\\
        300     &       0.21    &       17      &       0.016 & 0.6     &1.22e-04 &3.08e-03 &3.55e-03&2.46e-04\\
        1000    &       0.25    &       30      &       0.019 & 0.6     &1.22e-04 &3.21e-03 &3.93e-03&2.57e-04\\
        3000    &       0.28    &       43      &       0.022 & 0.6     & 6.10e-05 &2.12e-03 &2.75e-03 &1.70e-04\\
\hline
\label{para}
\end{tabular}
\vspace{0mm}
\parbox{0.9 \linewidth}{ Notes: $M_{\rm ecl}$ and $r_{\rm h}$ are the mass and the half-mass radius of the cluster.  $m_{\rm max}$ is the mass of the most massive star in the cluster. $\tau_{\rm g}$ and $\tau_{\rm d}$ are the decay time and the delay time during the gas expulsion process. 
The tree time step (-s), the changeover radius (-r), the neighbor searching radius (--r-search-min), and the multiple group radius (--r-bin) are the important parameters in \texttt{PeTar}.}
\end{table*}

\subsubsection{Fast gas expulsion}

The parameters for the simulation are summarized from previous works (i.e.,
\citealt{Kroupa2001-321,
Baumgardt2007-380,
Banerjee2012-746,
Banerjee2013-764,
Banerjee2014-787,
Banerjee2015-447,
Oh2015-805,
Oh2016-590,
Banerjee2017-597,
Brinkmann2017-600,
Oh2018-481,
Wang2019-484,
Pavlik2019-626,
Dinnbier2022-660,
Zhou2024rm}).
The previous simulations have already demonstrated the effectiveness and rationality of the parameter settings (see below). The influence of different parameter settings on simulation results and the discussion of the multidimensional parameter space can also be found in the works cited above. In this work, we mainly utilized the mature simulation recipe to interpret observational data. The model parameters of the simulations are listed in Table.\ref{para}. 

We computed three clusters with star masses of 300 M$_{\odot}$, 1000 M$_{\odot}$, and 3000 M$_{\odot}$ in the first 100 Myr.
The initial density profile of all clusters was the Plummer profile \citep{Aarseth1974-37, HeggieHut2003, Kroupa2008-760}, an appropriate choice since the molecular clouds’ filaments in which stars form have been found to have Plummer-like cross sections \citep{Malinen2012-544,Andre2022-667},
and open star clusters can also be described by the Plummer model \citep{Roeser2011-531,Roeser2019-627}.
Moreover, such a specific initial profile does not significantly affect the overall expansion rate of a cluster, as discussed in \citet{Banerjee2017-597}, 
which is primarily governed by the total stellar mass loss and the dynamical interactions occurring within the inner part of the cluster. 
The half-mass radius, $r_{h}$, of the cluster is given by the $r_{\rm h}-M_{\rm ecl}$ relation \citep{Marks2012-543}:
\begin{equation}
 \frac{r_{\rm h}}{{\rm pc}}=0.10_{-0.04}^{+0.07}\times\left(\frac{M_{\rm ecl}}{M_{\odot}}\right)^{0.13\pm0.04}\;.
 \label{rm}
\end{equation}
All clusters are fully mass segregated ($S$=1), with no fractalization, and in a state of virial equilibrium ($Q$=0.5). $S$ and $Q$ are the degree of mass segregation and the virial ratio of the cluster, respectively. More details can be found in \citet{Kupper2011-417} and the user manual for the \texttt{McLuster} code.
The initial segregated state is detected for young clusters and star-forming clumps/clouds \citep{Littlefair2003-345,Chen2007-134,Portegies2010-48,Kirk2011-727,Getman2014-787,Lane2016-833,
Alfaro2018-478,Plunkett2018-615, Pavlik2019-626, Nony2021-645, Zhang2022-936,
Xu2024-270}, but the degree of mass segregation is not clear. 
In simulations of the very young massive clusters R136 and NGC 3603 with gas expulsion by \citet{Banerjee2013-764}, 
mass segregation does not influence
the results. In \citet{Zhou2024rm}, we compared $S$=1 (fully mass segregated) and $S$=0.5 (partly mass segregated) and found similar results. We also discussed settings with and without fractalization in \citet{Zhou2024rm}; the results of the two are also consistent.
The initial mass functions of the clusters were chosen to be canonical \citep{Kroupa2001-322}, with the most massive star following the $m_{\rm max}-M_{\rm ecl}$ relation  of \citet{Weidner2013-434}:
\begin{equation}
\label{eq:mmaxmecl2}
y = a_0 + a_1 x + a_2  x^2 + a_3  x^3 ,
\end{equation}
where $y$ = $\log_{10}(m_\mathrm{max}/M_\odot)$, $x$ = $\log_{10}(M_\mathrm{ecl}/M_\odot)$, $a_0$ = -0.66, $a_1$ = 1.08, $a_2$ = -0.150, and $a_3$ = 0.0084. We assumed the clusters to be at solar metallicity \citep[i.e., $Z=0.02$;][]{von2016ApJ...816...13V}. 
The clusters travel along circular orbits within the Galaxy, positioned at a galactocentric distance of 8.5 kpc, and are moving at a speed of 220 km s$^{-1}$.

The initial binary setup follows the method described in \citet{Wang2019-484}.
All stars are initially in binaries, that is to say, $f_{\rm b}$=1, where $f_{\rm b}$ is the primordial binary fraction.
\citet{Kroupa1995-277-1491,Kroupa1995-277-1507} propose that stars with masses below a few $M_\odot$ are initially formed with a universal binary distribution function and that star clusters start with a $100\%$ binary fraction. Inverse dynamical population synthesis was employed to derive the initial distributions of binary periods and mass ratios. \cite{Belloni2017-471} introduce an updated model of pre-main-sequence eigenevolution, originally developed by \cite{Kroupa1995-277-1507}, to account for the observed correlations between the mass ratio, period, and eccentricity in short-period systems. For low-mass binaries, we adopted the formalism developed in \citet{Kroupa1995-277-1491,Kroupa1995-277-1507} and \citet{Belloni2017-471} to characterize the period, mass ratio, and  eccentricity distributions.
For high-mass binaries (OB stars with masses $>5$~M$_\odot$), we utilized the \citet{Sana2012-337} distribution, which is derived from O stars in OCs. The distribution functions of the period, mass ratio, and eccentricity are presented in \cite{Oh2015-805} and \cite{Belloni2017-471}. 

Accurately modeling gas removal from embedded clusters is challenging due to the complexity of radiation hydrodynamical processes, which involve uncertain and intricate physical mechanisms. 
To simplify the approach, we simulated the key dynamical effects of gas expulsion by applying a diluting, spherically symmetric external gravitational potential to a model cluster, following the method presented in \citet{Kroupa2001-321} and \citet{Banerjee2013-764}. 
This analytical approach is partially validated by \citet{Geyer2001-323}, who conducted comparison simulations using the smoothed particle hydrodynamics method to treat the gas. The hydrodynamics+N-body simulations in \citet{Farias2024-527} also confirm that the exponential decay model presented in Eq. \ref{eq:mdecay} generally provides a good description of gas removal driven by radiation and wind feedback.
Specifically, we used the potential
of the spherically symmetric, time-varying mass distribution:
\begin{eqnarray}
M_g(t)=& M_g(0) & t \leq \tau_d,\nonumber\\
M_g(t)=& M_g(0)\exp{\left(-\frac{(t-\tau_d)}{\tau_g}\right)} & t > \tau_d.
\label{eq:mdecay}
\end{eqnarray}
Here, $M_g(t)$ is the total mass of the gas; it is spatially distributed with the Plummer density distribution \citep{Kroupa2008-760} and starts depleting after a delay of $\tau_d$, and is totally depleted 
within a timescale of $\tau_g$. The Plummer radius of the gas distribution is kept time-invariant \citep{Kroupa2001-321}.
This assumption is an approximate model of the effective gas reduction within the cluster in the situation that gas is blown out while new gas is also accreting into the cluster along filaments such that the gas mass ends up being reduced with time but the radius of the gas distribution remains roughly constant. As discussed in \citet{Urquhart2022-510},
the mass and radius distributions of the ATLASGAL clumps at different evolutionary stages are quite comparable.
We used an average velocity of the outflowing gas of $v_g\approx10$ km s$^{-1}$, which is the typical sound speed in an HII region. This gives
$\tau_g=r_h(0)/v_g$,
where $r_h(0)$ is the initial half-mass radius of the cluster. As for the delay time, we took the representative value of $\tau_d\approx0.6$ Myr
\citep{Kroupa2001-321}, this being about the lifetime of the ultracompact HII region. 
As shown in \citet{Banerjee2013-764}, varying the delay time, $\tau_d$, primarily results in a temporal shift in the cluster’s rapid expansion phase, without significantly impacting its subsequent evolution for times greater than $\tau_d$. 
Protoclusters typically form in hub-filament systems \citep{Motte2018-56,Vazquez2019-490, Kumar2020-642,Zhou2022-514}, which are located in hub regions. Compared to the surrounding filamentary gas structures, the hub region, as the center of gravitational collapse, is usually more regular, as shown in \citet{Zhou2022-514,Zhou2024-686}. 
Thus, modeling a spherically symmetric mass distribution is appropriate (see Sect. \ref{sm} for more discussion). 

In this work, we assumed a SFE $\approx$ 0.33 as a benchmark (i.e., $M_{g}(0)$ = 2$M_{\rm ecl}(0))$. This value has been widely used in the simulations cited above and has been proven effective in reproducing the observational properties of stellar clusters. Such a SFE is also consistent with the value obtained from hydrodynamical calculations that include self-regulation \citep{Machida2012-421,Bate2014-437} and as well with observations of embedded systems in the solar neighborhood \citep{Lada2003-41,Megeath2016-151}.
In \citet{Zhou2024-688}, by comparing the mass functions of the ATLASGAL clumps and the identified embedded clusters, we found that a SFE of $\approx$ 0.33 is typical for the ATLASGAL clumps, which are also presented in Fig.\ref{compare}(b) here. 
In \citet{Zhou2024arXiv240809867Z}, assuming SFE = 0.33, it was shown that the total bolometric luminosity of synthetic embedded clusters can fit the observed bolometric luminosity of ATLASGAL clumps
with HII regions. 
In \citet{zhou2024starformationhistoriesstar}, we directly calculated the SFE of ATLASGAL clumps with HII regions and found a median value of $\approx$0.3.

\subsubsection{Timescale of gas free}\label{free}

Seven of the 17 MSFRs investigated in the MYStIX project are covered in the ATLASGAL survey \citep{Schuller2009-504}: the Lagoon nebula, NGC~6334, NGC~6357, the Eagle nebula, M~17, the Carina nebula, and the Trifid nebula.
As shown in Appendix \ref{msfr}, the ATLASGAL+\textit{Planck} or only ATLASGAL 870 $\mu$m data \citep{Csengeri2016-585} were used to trace the gas distribution around embedded or very young clusters. 
On these maps, clusters and clumps are well separated. 
However, NGC~6334 and M~17 still have abundant gas, which means they are younger than the other MSFRs. As shown in Fig. \ref{age}, the clusters in NGC~6334 and M~17 ("a") are indeed younger (the median value is $\approx$1.4 Myr) than those in other MSFRs ("b";\ the median value is $\approx$2.2 Myr). If embedded clusters in NGC~6334 and M~17 just finished the gas expulsion, then this implies an upper limit on the gas expulsion time of around 1.4 Myr.


 
\subsubsection{Slow and moderate gas expulsions}\label{sm}

Embedded clusters form in clumps. More massive clumps can produce more massive clusters, leading to stronger feedback and a higher gas expulsion velocity \citep{Dib2013-436}. 
There should be correlations between the feedback strength, the clump (or cluster) mass, and the gas expulsion velocity ($v_g$). And low-mass clusters should have a slower gas expulsion process compared with high-mass clusters. As shown in \citet{Pang2021-912}, low-mass clusters tend to agree with the simulations of slow gas expulsion. Except for the feedback strength, the SFE determines the total amount of the remaining gas, which also influences the timescale of gas expulsion. The gas expulsion process is driven by feedback, and the effectiveness of the feedback will depend on the geometric shape of the gas shell surrounding the embedded cluster  \citep{Wunsch2010-407,Rahner2017-470}. Therefore, a complex or nonspherical gas distribution would also change the timescale of gas expulsion.
In short, these parameter uncertainties can ultimately be incorporated into the timescale of gas expulsion. Thus, apart from the fast gas expulsion with the gas decay time, $\tau_g$, described above, we also simulated slow and moderate gas expulsions. 
As described in Sect. \ref{free}, the upper limit on the gas expulsion timescale may be $\approx$1.4 Myr. Therefore, for the slow gas expulsion, the gas decay time was set to 10$\tau_g$. 
The moderate gas expulsion is between the fast and slow gas expulsions. Considering the large parameter space between $\tau_g$ and 10$\tau_g$, we simulated two kinds of moderate gas expulsions: 2$\tau_g$ ("moderate2") and 5$\tau_g$ ("moderate1") . 
Figure \ref{gas} shows the evolution of gas mass under different gas expulsion modes over time.

\subsubsection{Star formation efficiency}\label{sfe}

The total amount of the residual gas not only affects the gas expulsion timescale, but it also significantly influences the strength of the gas potential. The strength of the external gas potential may have a considerable impact on the evolution of embedded star clusters. The total amount of residual gas is determined by the SFE of the clump. In the simulations described above, we assumed a SFE = 0.33 for all star clusters. 
In \citet{zhou2024starformationhistoriesstar}, we calculated the SFE of ATLASGAL clumps with HII regions and found a strong anticorrelation between
the SFE and the clump mass. Here, we directly derived the residual gas mass using Eq. 6 in \citet{zhou2024starformationhistoriesstar}, which encodes the variation in SFE with clump mass:
\begin{equation}
\mathrm{log}_{10} (M_{\mathrm{cl}}) = (1.02 \pm 0.02) \times \mathrm{log}_{10} (M_{\mathrm{ecl}}) + (0.52 \pm 0.05),
\end{equation}
and
\begin{equation}
M_{g}(0) = M_{\mathrm{cl}}- M_{\mathrm{ecl}},
\end{equation}
where $M_{\mathrm{cl}}$ is the clump mass. 


\subsubsection{Procedure}
The \texttt{McLuster} program \citep{Kupper2011-417} was used to set the initial configurations. 
The dynamical evolution of the model clusters was computed using the state-of-the-art \texttt{PeTar} code \citep{Wang2020-497}. 
\texttt{PeTar} employs well-tested analytical single and binary stellar evolution recipes (SSE/BSE)
\citep{Hurley2000-315,Hurley2002-329,Giacobbo2018-474,Tanikawa2020-495,Banerjee2020-639}. In \texttt{PeTar}, there are some important parameters: the tree time step (-s), the changeover radius (-r), the neighbor searching radius (--r-search-min), and the multiple group radius (--r-bin). 
The tree time step is a fixed time step used to calculate the long-range (particle-tree) force. The changeover region is the overlap shell between the long-range and short-range interaction. In particular, -s and -r significantly affect the accuracy and performance of the simulations. If -s and -r are not specified, \texttt{PeTar} determines them automatically, assuming that the input model is a spherical symmetric star cluster with a King- or a Plummer-like density profile. The neighbor searching radius is defined for each particle to identify neighboring candidates expected to be within the changeover region during the subsequent tree time step. Upon determining the changeover radius (-r), \texttt{PeTar} automatically computes the neighbor searching radius. The multiple group radius (--r-bin) is used as the criterion for selecting group members, which is automatically determined based on the changeover inner radius. In this work, all these parameters were automatically determined by \texttt{PeTar} (see Table \ref{para}). 

\subsection{Simulation results analysis}\label{ana}

In Fig. \ref{simu-s}, as expected, the mass loss of the clusters strongly depends on the gas expulsion modes. 
In the scenario of fast gas expulsion, gas is removed almost instantaneously, which strongly impacts the clusters. In contrast, with slow gas expulsion, the gas disperses over a timescale significantly longer than the cluster's crossing (dynamical) time, enabling the cluster to adapt to the changing gravitational potential and expand while maintaining near dynamical equilibrium, as discussed in \citet{Baumgardt2007-380}. This quasi-static (or adiabatic) evolution results in substantially less stellar mass loss, as shown in Fig. \ref{simu-s}. Therefore, it best fits the mass distribution of the observed OCs. However, it fails to match the radius distribution, as shown in Fig. \ref{fit0}(d1). 
For the fast gas expulsion, the clusters cannot survive more than 100 Myr. Similarly, in \citet{Kroupa2001-322} and \citet{Brinkmann2017-600},
the massive cluster containing many O stars most likely survives as an OC. Low-mass embedded clusters or groups dissolve quickly, in part due to the loss of their residual gas.
A lower SFE means more residual gas, leading to more intense expansion and mass loss of the cluster after gas expulsion, as shown in Fig. \ref{simu-s}(b). A lower SFE is equivalent to a shorter gas expulsion timescale. 

For high-mass clusters, such as 3000 M$_{\odot}$, $\tau_g$ and 2$\tau_g$ will give significantly different results. Thus, the simulations are very sensitive to the gas decay time. We need to carefully constrain this parameter from observations in future work. Since $\tau_g=r_h(0)/v_g$, with $r_h(0)$ being constrained by Eq. \ref{rm}, we can focus on the measurement of the gas expulsion velocity ($v_g$) in observations, especially the correlation between $v_g$ and the clump (or cluster) mass, as discussed in Sect. \ref{sm}.

In Fig. \ref{fit0},
fast gas expulsion is required to explain the radius distribution of the observed OCs, but to account for the mass distribution, slow gas expulsion is necessary. This contradiction makes both fast and slow gas expulsions inappropriate. To reconcile this contradiction, the gas expulsion should be somewhere between fast and slow. In terms of both mass and radius, "moderate2" can best fit the observed OCs, and the main contribution is from the high-mass cluster (3000 M$_{\odot}$).
To explain the mass and radius distributions of the observed OCs, initial embedded clusters with masses higher than 3000 M$_{\odot}$ are necessary. However, in Fig. \ref{compare}, the upper limit of the embedded cluster sample is less than 1000 M$_{\odot}$, and only a few clumps have a mass higher than 3000 M$_{\odot}$. For an OC, its progenitor should have a significantly higher mass. Thus, the currently observed clumps cannot be the "direct" precursors of the currently observed OCs. 

\section{Discussion}

There may be two evolutionary tracks from embedded clusters to OCs: a single embedded cluster evolution scenario and a multiple embedded clusters merger (post-gas expulsion coalescence) scenario. 
In \citet{Zhou2024-688}, we used the MST method to measure the separations between clusters and gas clumps in each MSFR shown in Appendix \ref{msfr}. The separations between clusters, between clumps, and between clusters and clumps are comparable, which indicates that the evolution from clump to embedded cluster proceeds in isolation and locally, and does not significantly affect the surrounding objects. Using {\it Gaia} DR2 data, \citet{Kuhn2019-870} investigated the kinematics of subclusters identified in \citet{Kuhn2014-787} for the same MSFRs and found no evidence that these groups are merging. Therefore, in this work, we only focused on the first scenario.
However, as presented above, the masses of current clumps or embedded clusters in the Milky Way are too low to have evolved into the observed OCs.

For the analysis in this work, we assumed a constant star formation history for the Milky Way. If its  star formation history is instead burst-like and time-dependent, the currently observed OCs may be remnants of massive clumps from the past. The epoch of peak star formation rate density is thought to lie between redshifts of 2 and 3 \citep{Madau1998-498,Hopkins2006-651,Bouwens2011-737,Moster2013-428}. {\it Gaia} DR2 reveals a star formation burst in the Galactic disk 2-3 Gyr ago \citep{Mor2019-624}. As an inference, the peak period of star formation in the Milky Way has passed, and most of the massive clumps have evolved into OCs or older clusters. At present, only relatively low-mass clumps remain. This possibility is worth exploring further in future work. 

Another explanation is that these massive and relatively old OCs come from post-gas expulsion coalescence of multiple embedded clusters. As shown in Appendix \ref{msfr}, the embedded clusters or the very young clusters  are very near to each other in each MSFR. There is a very real possibility that they will merge after expansion. One molecular cloud can produce multiple embedded clusters, which can merge into larger clusters. This scenario is consistent with the finding in Sect. \ref{sepa}, that most molecular clouds can only form one OC. 
In a forthcoming work (Zhou et. al. in preparation), we will simulate the post-gas expulsion coalescence of subclusters in each MSFR, with the initial conditions of the simulations directly derived from the observation. 
There is now extensive literature arguing that star clusters form from mergers between embedded clusters, both from simulations \citep{Fujii2012-753,Howard2018-2,Sills2018-477,Fujii2022-514,Dobbs2022-517,Cournoyer2023-521,Reina2024arXiv} and observations \citep{Sabbi2012-754,Pang2022-931,Della2023-674,Fahrion2024-681}. A comparison with these works will also be presented in the forthcoming work.

In the current work, the initial density profile of all clusters is the Plummer profile. However, observations of embedded clusters show clusters with high ellipticities \citep{Kuhn2014-787,GetmanKV2018Ysci}.
Due to the limitations of the method used here, we are currently unable to discuss the impact of the cluster's ellipticity on its subsequent evolution. For now, we have only performed N-body simulations. In future work, it is necessary to compare them with more comprehensive hydrodynamics+N-body simulations, such as those presented in \citet{Fujii2021-73}, \citet{Rieder2021-509}, and \citet{Polak2024arXiv240512286P}, to further test and constrain the initial conditions used in the N-body simulations presented in this work.

\section{Conclusion}

We collected samples of Galactic clumps, embedded clusters, and OCs to compare their physical properties.  Overall, the radii of OCs are significantly larger than those of embedded clusters and clumps. The radii of embedded clusters are larger than those of clumps, which may be
due to the expansion of embedded clusters.
The mass distribution of OCs covers a significantly larger mass range than that of embedded clusters, by about one order of magnitude.
Given the current mass distribution of clumps in the Milky Way, it is unlikely that a single clump evolving into a single embedded cluster, and subsequently into a single OC, could account for the observed OCs with old ages and high masses, which is also supported by N-body simulations of individual embedded clusters.  

We used standard N-body simulations as in previous works to interpret the physical parameters of the observed OCs. 
The mass loss of the simulated clusters strongly depends on the gas expulsion modes.  For the fast gas expulsion, the clusters ($\leq$3000 M$_{\odot}$) cannot survive more than 100 Myr.
Fast gas expulsion is required to explain the radius distribution of the observed OCs, but to account for the mass distribution, slow gas expulsion is necessary; the real gas expulsion speed should be somewhere in between. In any case, to explain the mass and radius distributions of the observed OCs, initial embedded clusters with masses higher than 3000 M$_{\odot}$ are necessary. However, in Fig. \ref{compare}, the upper limit of the embedded cluster sample is less than 1000 M$_{\odot}$, and only a few clumps have a mass higher than 3000 M$_{\odot}$. The progenitor of an OC should have a significantly higher mass. Thus, the currently observed clumps cannot be the "direct" precursors of the currently observed OCs. 

If the  star formation history of the Milky Way is burst-like and time-dependent, the currently observed OCs may be remnants of massive clumps from the past.  At present, only relatively low-mass clumps remain. 
Another explanation is that those massive and relatively old OCs come from post-gas expulsion coalescence of multiple embedded clusters.
We compared  the separation of OCs and the typical size of molecular clouds in the Milky Way and find that most molecular clouds can only form one OC. Each molecular cloud typically includes many embedded clusters; 
the typical separation between them is $\approx$1 pc \citep{Zhou2024-688}. Thus, after expansion, they should be able to undergo mergers.
Further study is necessary to distinguish between the different scenarios.


\begin{acknowledgements}
Thanks the referee for providing detailed and constructive review comments, which have helped to improve and clarify this work.
\end{acknowledgements}

\bibliographystyle{aa} 
\bibliography{ref}


\appendix


\section{Massive star-forming regions}\label{msfr}
\begin{figure*}[htbp!]
\centering
\includegraphics[width=0.75\textwidth]{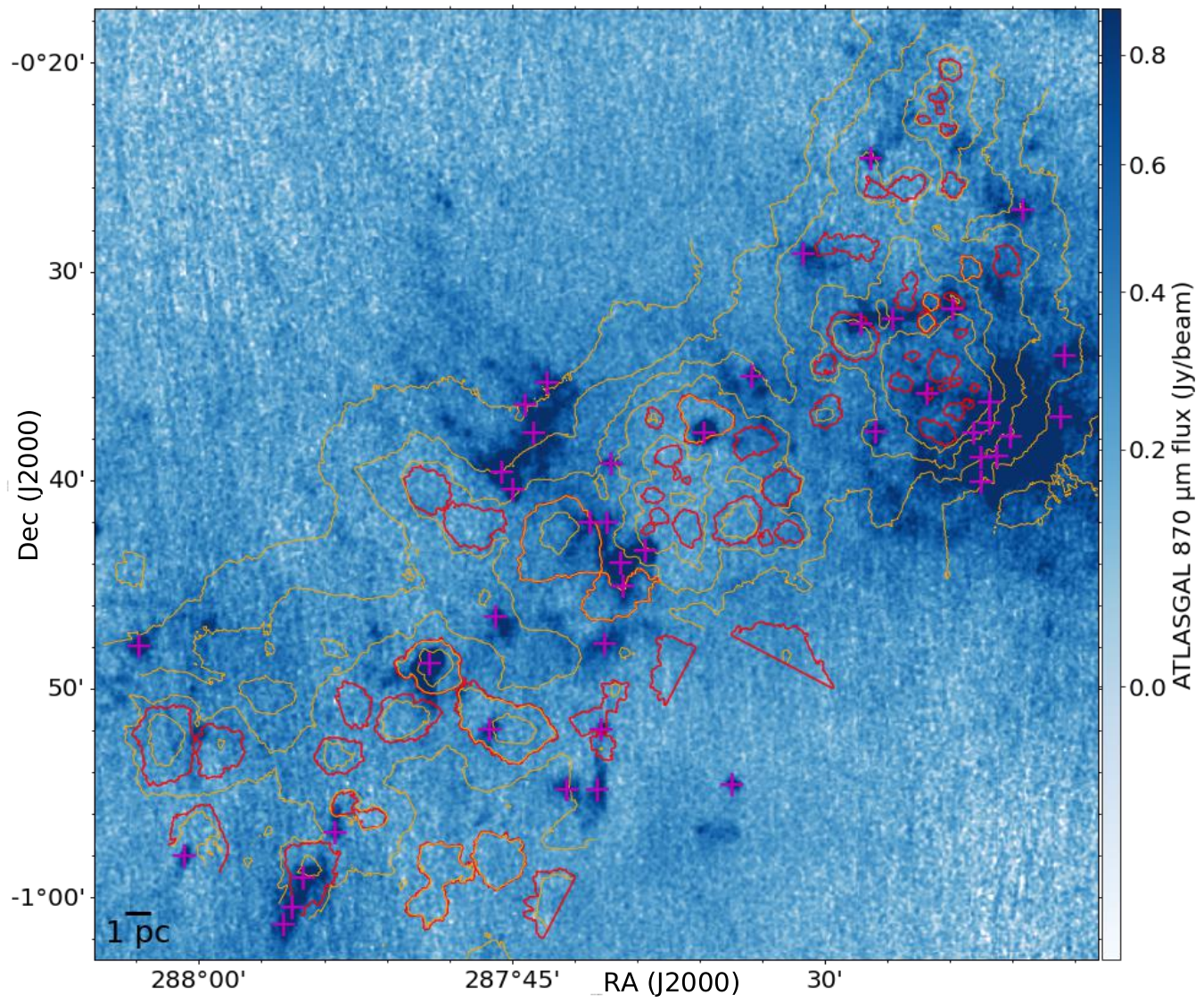}
\caption{Carina. The background is the ATLASGAL 870 $\mu$m emission. Red contours are the subclusters identified by the dendrogram algorithm based on the surface density distribution of stars from the MYStIX project in \citet{Zhou2024-688}. Cyan "+" symbols mark the ATLASGAL clumps.  The background density of the stars' surface density map is $\Sigma_{\rm rms}$. Orange contours show [2, 4, 8, 16, 32] $\times \Sigma_{\rm rms}$.
}
\label{Carina}
\end{figure*}

\begin{figure}[htbp!]
\centering
\includegraphics[width=0.5\textwidth]{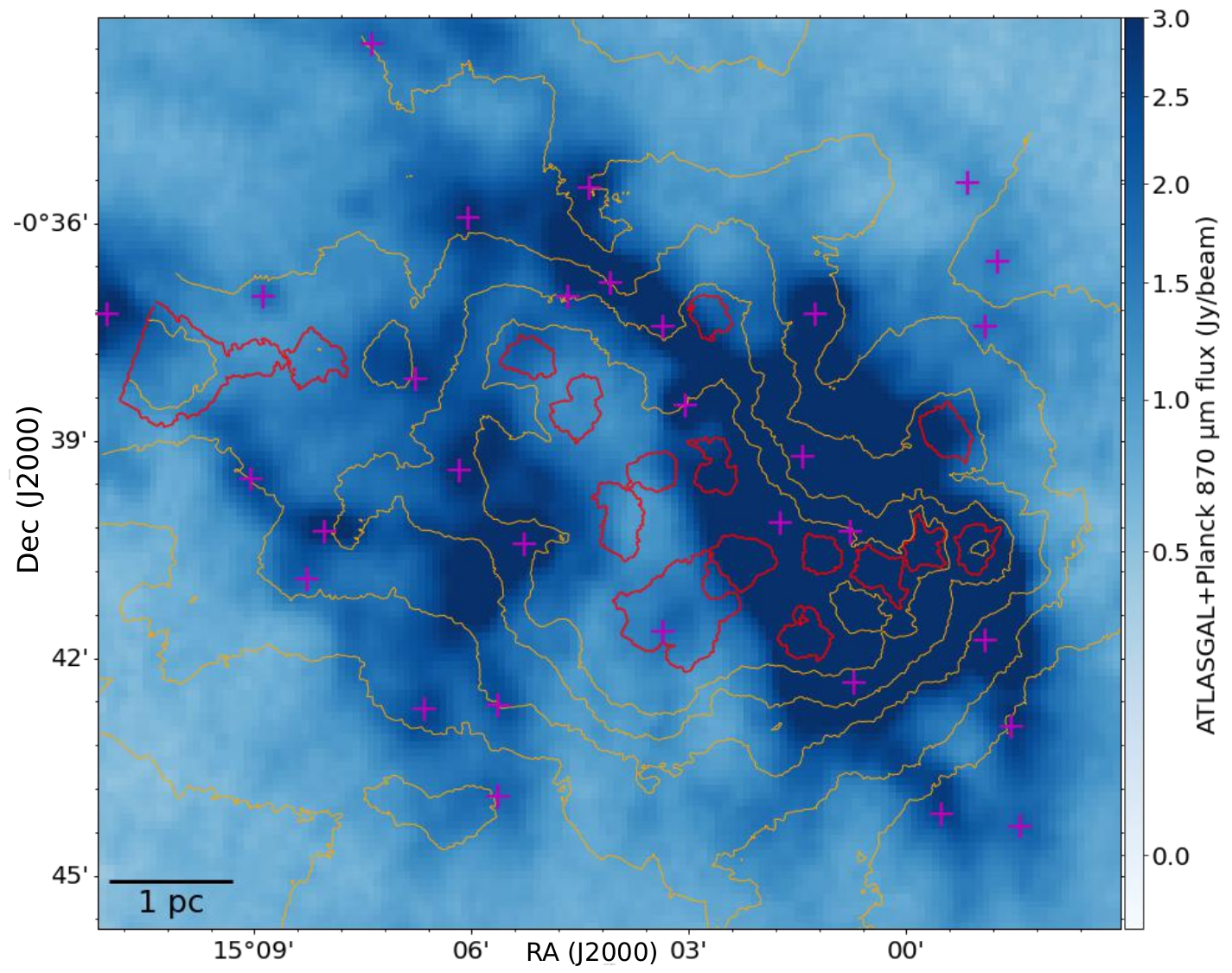}
\caption{Same as Fig. \ref{Carina} but for M17. }
\label{M17}
\end{figure}

\begin{figure}[htbp!]
\centering
\includegraphics[width=0.5\textwidth]{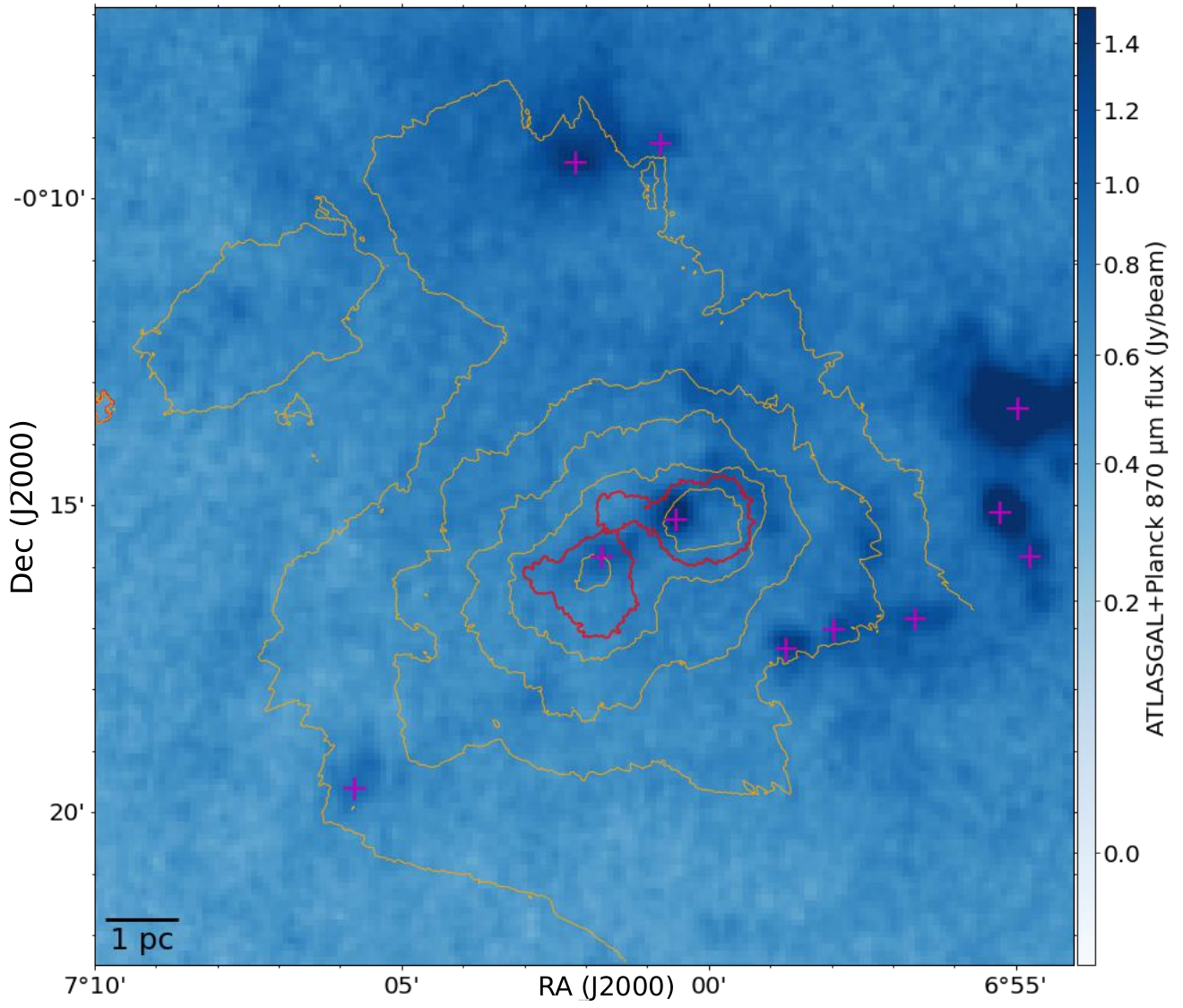}
\caption{Same as Fig. \ref{Carina} but for Trifid. }
\label{Trifid}
\end{figure}

\begin{figure}[htbp!]
\centering
\includegraphics[width=0.5\textwidth]{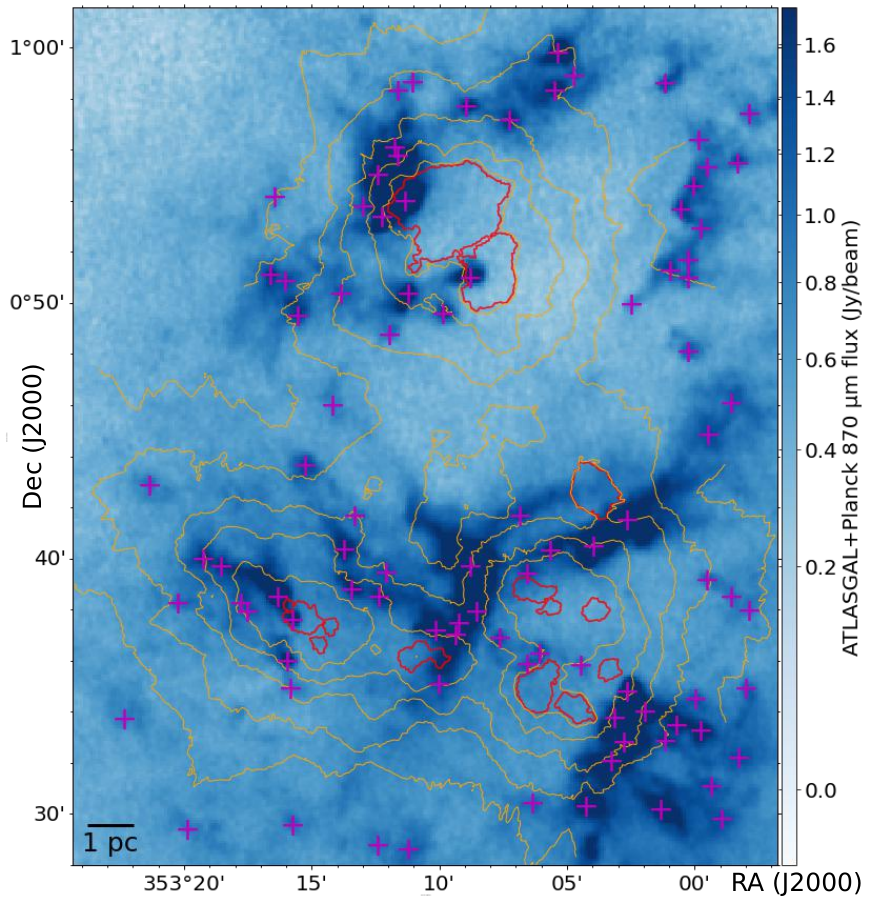}
\caption{Same as Fig. \ref{Carina} but for NGC6357. }
\label{NGC6357}
\end{figure}

\begin{figure}[htbp!]
\centering
\includegraphics[width=0.5\textwidth]{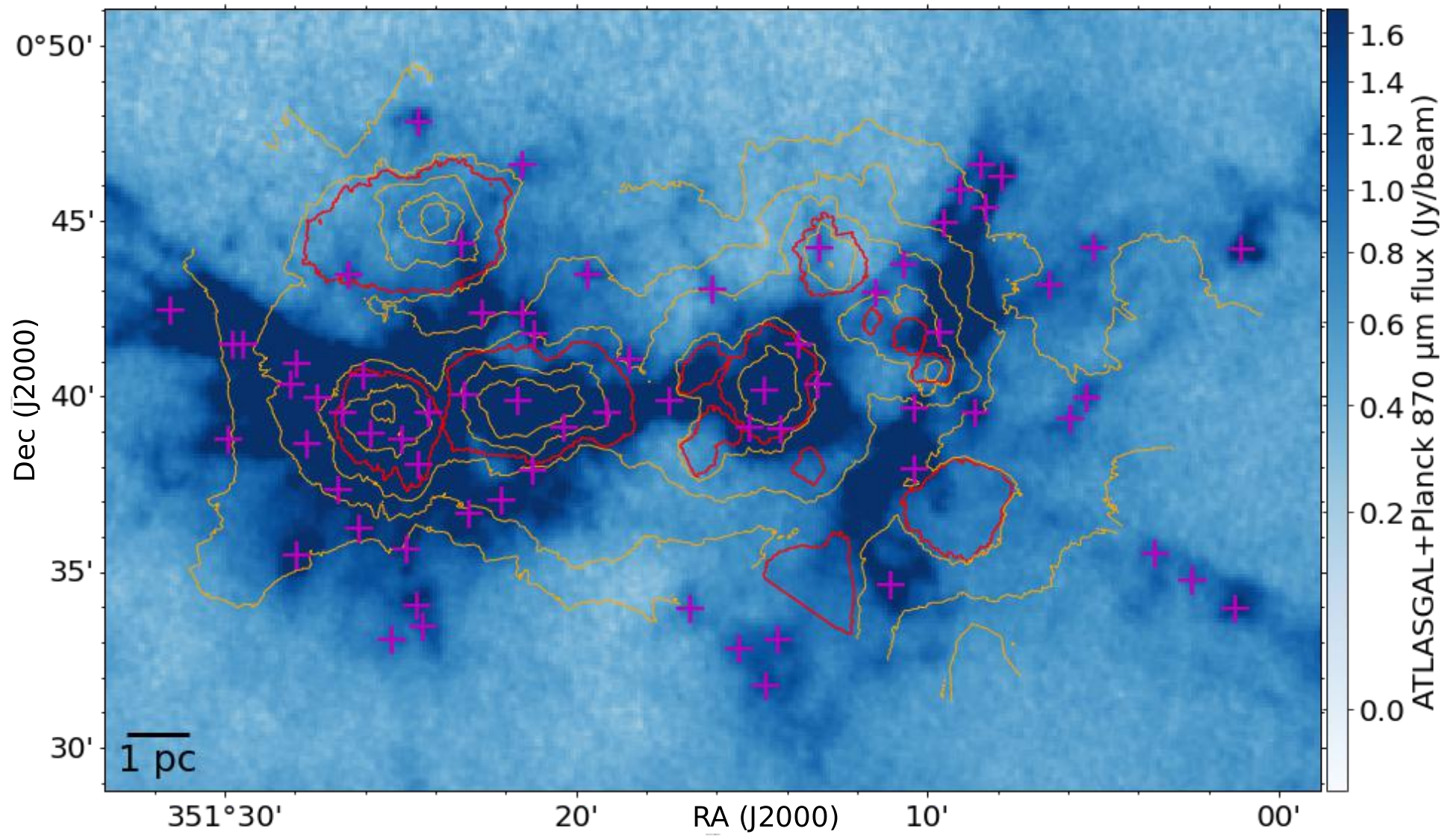}
\caption{Same as Fig. \ref{Carina} but for  NGC6334.}
\label{NGC6334}
\end{figure}

\begin{figure}[htbp!]
\centering
\includegraphics[width=0.5\textwidth]{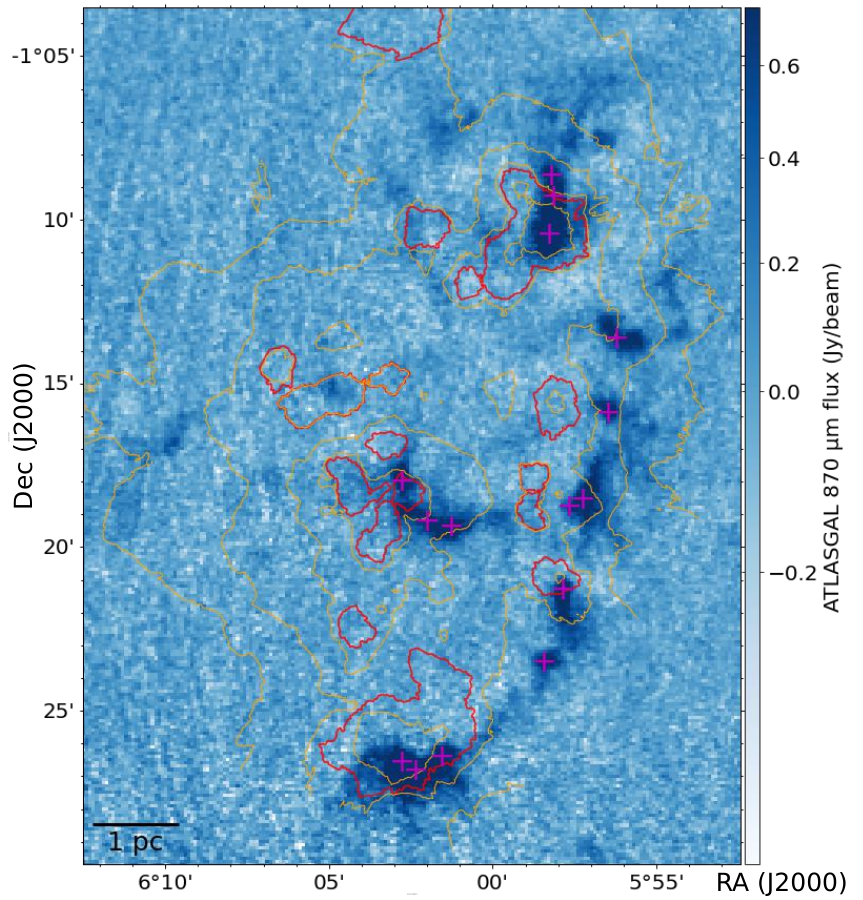}
\caption{Same as Fig. \ref{Carina} but for Lagoon. }
\label{Lagoon}
\end{figure}

\begin{figure}[htbp!]
\centering
\includegraphics[width=0.5\textwidth]{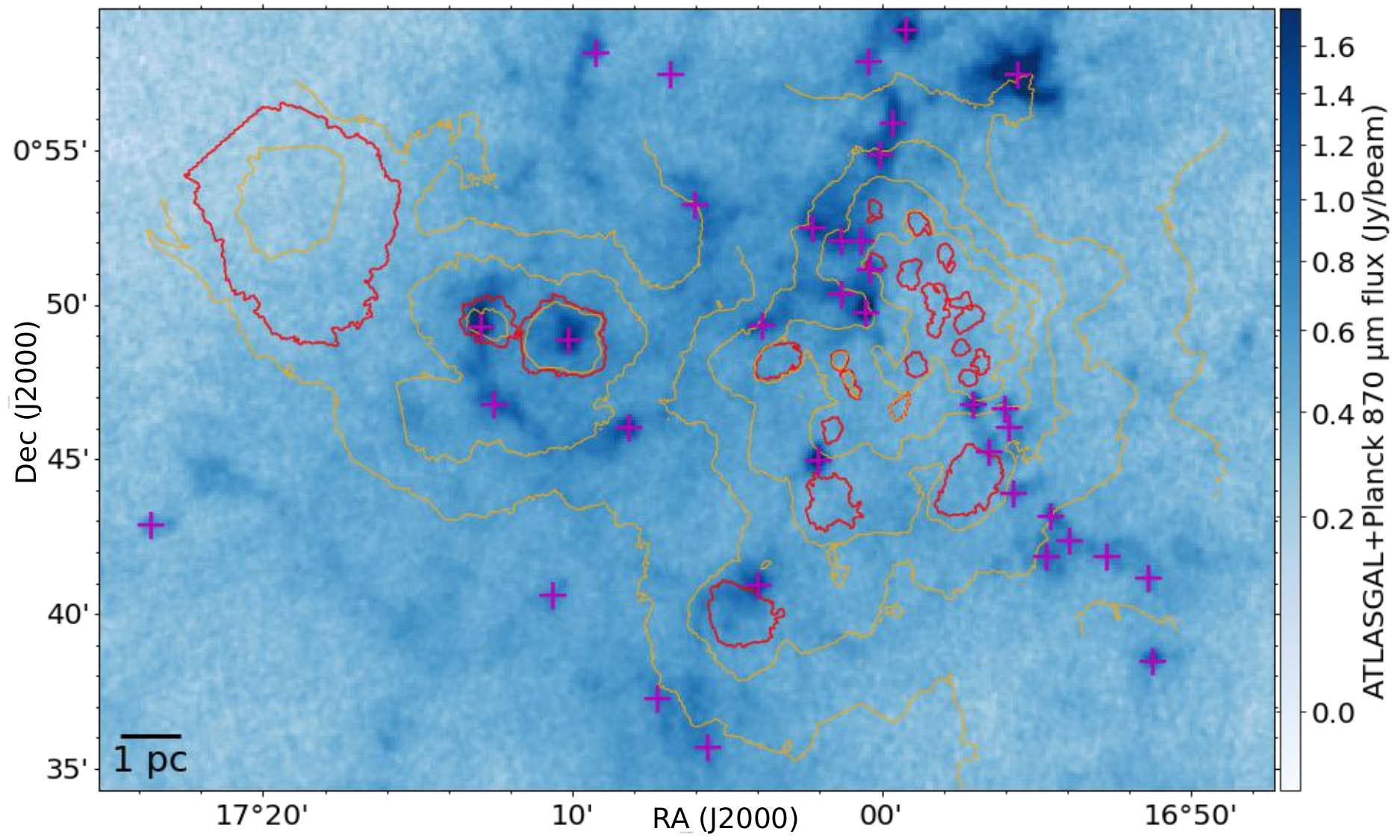}
\caption{Same as Fig. \ref{Carina} but for Eagle. }
\label{Eagle}
\end{figure}

\end{document}